\documentstyle[11pt,dunk2001_asp,twoside,psfig]{article}
\begin{document}

\def\emphasize#1{{\sl#1\/}}
\def\arg#1{{\it#1\/}}
\let\prog=\arg

\def\edcomment#1{\iffalse\marginpar{\raggedright\sl#1\/}\else\relax\fi}
\marginparwidth 1.25in
\marginparsep .125in
\marginparpush .25in
\reversemarginpar

\def\kms {\ km s$^{-1}$}
\def\msol{\ifmmode {\>M_\odot}\else {$M_\odot$}\fi}
\def\cmsq{\ifmmode {\>{\rm\ cm}^2}\else {cm$^2$}\fi}
\def\psqcm{\ifmmode {\>{\rm cm}^{-2}}\else {cm$^{-2}$}\fi}
\def\psqpc{\ifmmode {\>{\rm pc}^{-2}}\else {pc$^{-2}$}\fi}
\def\pcsq{\ifmmode {\>{\rm\ pc}^2}\else {pc$^2$}\fi}
\def\Tkev{\ifmmode{T_{\rm kev}}\else {$T_{\rm keV}$}\fi}
\def\hubunits{\ifmmode {\>{\rm km\ s^{-1}\ Mpc^{-1}}}\else {km
s$^{-1}$ Mpc$^{-1}$}\fi}
\def\gta{\;\lower 0.5ex\hbox{$\buildrel > \over \sim\ $}}
\def\lta{\;\lower 0.5ex\hbox{$\buildrel < \over \sim\ $}}

\def\phiIV{\ifmmode{\varphi_4}\else {$\varphi_4$}\fi}
\def\phiI{\ifmmode{\varphi_i}\else {$\varphi_i$}\fi}

\def\be{\begin{equation}}
\def\ee{\end{equation}}
\def\bea{\begin{eqnarray}}
\def\eea{\end{eqnarray}}
\def\beas{\begin{eqnarray*}}
\def\eeas{\end{eqnarray*}}
\def\gtrapprox{\;\lower 0.5ex\hbox{$\buildrel >\over \sim\ $}}
\def\lessapprox{\;\lower 0.5ex\hbox{$\buildrel < \over \sim\ $}}
\def\deg   {$^\circ$}
\def\Ftwo  {$F_{-21}$}
\def\Pcos  {$\Phi^0$}
\def\Jtwo  {$J_{-21}$}
\def\Fcos  {$F_{-21}^0$}
\def\Jcos  {$J_{-21}^0$}
\def\Em    {${\cal E}_m$}
\def\ALL   {A_{\scriptscriptstyle LL}}
\def\JLL   {J_{\scriptscriptstyle LL}}
\def\nuLL  {\nu_{\scriptscriptstyle LL}}
\def\sigLL {\sigma_{\scriptscriptstyle LL}}
\def\tauLL {\ifmmode{\tau_{\scriptscriptstyle LL}}\else 
           {$\tau_{\scriptscriptstyle LL}$}\fi}
\def\nuOB  {\nu_{\scriptscriptstyle {\rm OB}}}
\def\aB    {\alpha_{\scriptscriptstyle B}}
\def\nH    {n_{\scriptscriptstyle H}}
\def\Em{\ifmmode{{\rm E}_m}\else {{\rm E}$_m$}\fi}
\def\NH{\ifmmode{{\rm N}_{\scriptscriptstyle\rm H}}\else {{\rm N}$_{\scriptscriptstyle\rm H}$}\fi}

\def\Ha    {H$\alpha$}
\def\Hb    {H$\beta$}
\def\HI    {H${\scriptstyle\rm I}$}
\def\HII    {H${\scriptstyle\rm II}$}
\def\eg    {{\it e.g.}}
\def\ie    {{\it i.e.}}
\def\cf    {{\it cf. }}
\def\qv    {{\it q.v. }}
\def\etal  {\ et al.}
\def\kms{\ifmmode {\>{\rm\ km\ s}^{-1}}\else {\ km s$^{-1}$}\fi}

\def\Em{\ifmmode{{\cal E}_m}\else {{\cal E}$_m$}\fi}
\def\Dm{\ifmmode{{\cal D}_m}\else {{\cal D}$_m$}\fi}
\def\fesc{\ifmmode{\hat{f}_{\rm esc}}\else {$\hat{f}_{\rm esc}$}\fi}
\def\fescs{\ifmmode{f_{\rm esc}}\else {$f_{\rm esc}$}\fi}
\def\rsolar{\ifmmode{r_\odot}\else {$r_\odot$}\fi}
\def\emunit{\ifmmode{{\rm cm}^{-6}{\rm\ pc}}\else {
cm$^{-6}$ pc}\fi}
\def\intensity{\ifmmode{{\rm erg\ cm}^{-2}{\rm\ s}^{-1}
      {\rm\ Hz}^{-1}{\rm\ sr}^{-1}}
      \else {erg cm$^{-2}$ s$^{-1}$ Hz$^{-1}$ sr$^{-1}$}\fi}
\def\flux{\ifmmode{{\rm erg\ cm}^{-2}{\rm\ s}^{-1}}\else {erg
cm$^{-2}$ s$^{-1}$}\fi}
\def\fluxdensity{\ifmmode{{\rm erg\ cm^{-2}\ s^{-1}\ Hz^{-1}}}\else {erg
cm$^{-2}$ s$^{-1}$ Hz$^{-1}$}\fi}
\def\phoflux{\ifmmode{{\rm phot\ cm}^{-2}{\rm\ s}^{-1}}\else {phot
cm$^{-2}$ s$^{-1}$}\fi}
\def\phorate{\ifmmode{{\rm phot\ s}^{-1}}\else {phot s$^{-1}$}\fi}

\def\apj{{\it Ap.J.~}}
\def\apjs{{\it Ap.J.Suppl.~}}
\def\apss{{\it Astrophys.Sp.Science~}}
\def\aj{{\it Astron.J.~}}
\def\aph{{\it astro-ph}}
\def\mn{{\it MNRAS~}}
\def\araa{{\it Annu.Rev.Astron.Astrophys.~}}
\def\pasp{{\it PASP.~}}
\def\aaa{{\it Astron.Astrophys.~}}
\def\aaas{{\it Astron.Astrophys.Suppl.~}}
\def\astroph{{\it astro-ph}}
\def\rmp{{\it Rev.Mod.Phys.~}}

\setcounter{page}{373}

\title{\Large\bf The Dynamics, Structure and History of Galaxies: Workshop 
Summary}

\author{Joss Bland-Hawthorn} 
\affil{Anglo-Australian Observatory, 167 Vimiera Road, Eastwood, NSW 2122, Australia}

\section{Introduction}

As far as anyone knew, we were gathered at the resplendent Dunk Island
on the Barrier Reef to celebrate Ken Freeman's 60th birthday.  But what
we got was a well presented cross-section of key topics in Galactic and
extragalactic astronomy. Many of the speakers were former students of Ken's
and the quality of the talks reflected this fact.  If there was an
overarching theme to the meeting, it reflected Ken's wish to understand
``the chain of events that occurred during the formation of the Galaxy",
to quote from his Annual Reviews article in 1987.
The four day workshop was divided into four parts:
galaxy structure and dynamics, haloes and environments, HI and dark
matter, galaxy formation and evolution.

\section{Galaxy Structure and Dynamics}

Van der Kruit began the meeting with a bold statement: Ken Freeman's
celebrated paper in 1970 remains the most highly cited in {\it
Astrophysical Journal} of that year, and falls within the Top Ten of
the most cited papers in galactic astronomy. 
He went on to review what we have learnt over the intervening years about
galaxy disks, concluding that the central surface brightness $\mu_o$
given by ``Freeman's law'' -- $\mu_o$ is approximately constant for
spiral disks -- is really an upper limit on $\mu_o$.  It is interesting
to note that, in the original paper, Ken makes no statement that would
constitute a law, although he does explore the theoretical consequences
of constant $\mu_o$. But the ``Freeman disk'' continues to play an
important role in observational and theoretical studies.  

The remarkable fact of the existence of the Tully-Fisher relation 
continues to puzzle.
De~Blok outlined the intimate connection of the Tully-Fisher conundrum
with galaxy formation and evolution, a theme picked up by Bosma in
his review of new HI data of late-type spirals. Bosma retraced the
history of the dark matter problem and emphasized the importance of
Ken's 1970 paper in stressing the mass discrepancy in galaxy dynamics,
long before the spate of `dark halo' theory papers starting in 
1973.  Throughout the 1970s, the observational case for dark matter
became more compelling largely due to HI studies of spirals by Bosma,
Roberts and others. From new observations of late-type spirals, Bosma
finds the HI rotation curves are much better fit with an isothermal
halo, rather than the NFW profile from CDM simulations.

Kormendy demonstrated the now-famous relation between the black hole
mass and the depth of the potential well as measured by the velocity
dispersion. There is a clear link between bulges and black holes which
was probably established in the early universe during the major epoch
of bulge growth.  M33, which has no bulge, appears to have no black
hole to an upper mass limit of 1000 solar masses. Sadler showed
evidence for a radio-power/black hole mass relation which may allow us
to track the evolution of black hole mass with redshift, something that
is presently difficult to do for bulges.

Planetary nebulae are proving to be excellent probes of the outermost
halo.  Ford showed the kinematics of many hundreds of planetary nebulae
out to 6 scale lengths in Cen A. The bisymmetric kinematic signature is
strongly suggestive of a tumbling or triaxial potential. An obvious
application would be to look for planetary nebulae in the faint stellar
streams which are seen to encircle a number of nearby galaxies,
including both M31 and the Galaxy.

Bureau, on behalf of the SAURON team, presented integral field results
for a large sample of elliptical, lenticular and spiral galaxies divided
between clusters and the field. Fully one third of these show evidence
of nuclear sub-structure. Some show evidence for triaxility, including
nuclear bars as judged from Athanassoula's bar simulations.  The group
emphasizes caution in the analysis and derivation of black hole masses
from longslit measurements. 

Kalnajs
entertained us with on-line simulations of Freeman bars and showed
that they come in three flavours; slow, fast and pointy. 
Block presented a new physical classification of galaxies based on
near-IR imaging. The method uses Elmegreen's proposal to estimate
bar torque from the projected mass distribution. The scheme is based on
the Fourier h$_2$ component of the IR light and the degree of arm
winding.  Block finds that ``once you penetrate the mask,'' many
multi-arm spirals are found to be grand design 2-arm spirals in the
near IR. There appears to be a major overlap in bar torque between
normal and barred spirals.  This resonated with Gerhard who presented
the crucial facts for a bar in the Galaxy. The predicted gas flow
patterns reveal four arms -- ``four arms good, two arms bad" -- which
appear to agree with the tangent points for the spiral arms defined by
HII regions. The COBE/DIRBE maps appear to require at least two spiral
arms, but beyond that, not much can be deduced.

\section{Haloes and environments}

One of the most active research areas of astrophysics, especially so in
Australia, concerns the chemical enrichment of the Galactic halo. From
Norris' review, it was clear that we have barely begun to tap the
information available from starlight. One of the most interesting
results is the dramatic fall off in the number of stars with [Fe/H] $<$
--4 with only a few currently known. A simple one-zone model of galactic
chemical enrichment predicts that the number of stars should decline by
a factor of ten for every decrease of 1 dex in [Fe/H]. Why this is not
the case below [Fe/H] $<$ --4 appears to be saying something important
about the nature of the first stars in the universe.

Some of the heavy r- and s-process elements in metal-poor stars display
an increasingly large scatter as [Fe/H] decreases. This remarkable
trend is thought to reflect incomplete mixing of the ISM in the early
Galaxy. In his talk, Bland-Hawthorn asked to what extent we could unravel 
disk formation from unique chemical signatures back to the last major 
epoch of dissipation. 

Dwarf galaxies continue to present puzzles. Some like Sculptor and
Fornax have low M/L values, while others like Sextans, Draco, Carina
and Ursa Minor have huge overdensities of dark matter. It seems likely
that some of the observed dwarfs have undergone dynamical stripping by
the Galaxy.  But Da Costa showed that these objects apparently cannot have
contributed a large fraction of the halo stars: while the iron peak
element abundances are much like field stars, the [$\alpha$/Fe] ratio
is substantially lower.

In the words of Mateo, the Local Group is a dangerous place for a
dwarf. The tidal radius of a dwarf with mass $m_7$ (in units of 10$^7$
M$_\odot$) at a distance $r_{100}$ (in units of 100 kpc) from the centre
of the Galaxy is roughly $r_d \sim r_{100} \root 3 \of {m_7}$. The
velocity dispersion of the dwarf is $\sigma_d \sim 6 \root 3 \of {m_7}
/ \sqrt{r_{100}}$ (in units of km s$^{-1}$). We can approximate the
timescale for dynamically stripping a dwarf galaxy as $\tau_d \sim
2\sqrt{3}(r_d/\sigma_d)$. A dwarf galaxy that ventures within 100 kpc 
of the two great rajahs of the Local Group can be stripped in just
a few gigayears. In the coming decade, it is clear that detailed heavy
element abundances (beyond those shown by Da Costa) will provide a
clearer picture as to what fraction of the halo stars formed {\it in
situ} rather than in satellite galaxies.

Future astrometric missions will no doubt find stellar streams
associated with open clusters, globular clusters, and events unseen.
Maybe all of the halo, much of the disk and some fraction of the bulge
is made up of streams and sub-structure.  Morrison spoke of the
Spaghetti Survey which should be sensitive to such events out to a
radius of 100 kpc. Clumping in the halo is also the main science goal
of the YSTAR project, as presented by Byun.  

De Zeeuw showed that
the Gaia mission at the end of the decade will be able to measure the
parallax and proper motions of 1.3 billion stars down to the level of
micro-arcseconds. At these levels, the gravity field of Jupiter is
detectable over the entire sky! Andrew Murray, author of an authorative
text on vectorial astrometry (``which dares not to venture below
milli-arcsecond levels''), once remarked that it would take a very
bright individual a lifetime to get to the bottom of all
relativistic phenomena at micro-arcsecond levels. This of course 
provides a rich data set in its own right.

King presented his analysis, in collaboration with Jay Andersen, of
Hubble Space Telescope data on M4 which allows for a near-perfect
separation of the cluster main sequence from the field stars from the
proper motions.  The dispersion in the main sequence becomes almost
immeasurable after the line-of-sight objects are rejected.  This is
reminiscent of recent work on open clusters which used the Hipparcos
database to weed out objects not associated with the clusters. Quillen
finds that the dispersion in [Fe/H] is immeasurable once line-of-sight
objects are rejected.

\section{HI and Dark matter}

While the ``successes'' of CDM on linear scales are trumpeted on
a routine basis, this cannot be said for CDM predictions of the
non-linear domain. This problem is particularly acute in regions of
large density contrast. CDM simulations overpredict the cuspiness of
galaxies, the dark matter fraction within the Solar Circle, the random
motions in the Local Supercluster, and the number satellites on the
scale of the Local Group.  Several speakers (e.g.\ Fall) noted that
relaxing the cuspiness of CDM profiles makes everyone's life easier.
But until this issue is resolved, interest in modified gravity laws
will continue, even in the absence of an inertial framework. Sellwood
showed the extent to which the MOND prescription can be made to explain
a diverse array of data, particularly the fact that dark haloes appear
to mimic the luminous components, e.g.\ squashed stellar systems have
squashed dark haloes.

Sadler, Carignan and Putman presented some of the major
results to derive from the southern sky HIPASS survey. It seems that
the Ly-alpha absorption systems arise primarily in the low mass HI
ellipticals are detected in HI in the HIPASS survey, and some of these
structures stretch to vast distances. Gibson showed that most
high-velocity clouds are likely to fall within 100 kpc of the Galaxy,
although Putman finds that some fraction of these might extend further
still. This, however, assumes that some compact clouds are physically
associated with dark matter haloes. It is therefore noteworthy that
Carignan claims evidence of HI associated with several dwarf
spheroidals. In the Galaxy, Allen presented evidence for the
association of HI with molecular gas which is being dissociated by the
soft UV radiation field. This link can be seen over spatial scales 
which cover many orders of magnitude (1 pc to 10 kpc).  Calzetti discussed
how detailed observations of local galaxies can yield insight into the 
multiple processes associated with the evolution of starbursts.

The detailed connection of baryons to dark matter continues to elude.
We should prepare ourselves for the possibility that there is more to
the dark matter universe than we might suppose.  CDM appears to predict
that the dark matter universe is largely uninteresting. But the
galaxies we know and love are the light cusps within the dark matter
potential, much as cD galaxies are to galaxy clusters. Dark matter
properties may well set up the visible universe to the extent that local
dark matter properties correlate with Hubble type and luminosity.

\section{Galaxy formation and evolution}

Fall showed that the salient features of the Fall \& Efstathiou model of
disk formation survive within the CDM hierarchy. But, as Weil et al.\ (1998,
MNRAS, 300, 773)
point out, this does require that most of the gas in small systems
stays hot long enough to settle to the disk when most of the accretion
activity has passed.  Fall and Silk both emphasized the importance of
viscosity and its relation to star-formation timescales. In the Lin \&
Pringle picture, this viscosity is maintained by the disk-halo
interaction and spiral arms.

Jerjen demonstrated that the Surface Brightness Fluctuation method can be 
used to determine distances to dwarf elliptical galaxies.  This then
opens the way for mapping  in a simple and effective way the detailed distribution of galaxies within $\sim$10 Mpc.  This may allow us to understand 
how large scale structures within this volume, such as the Supergalactic 
Plane, arise.  
Student {\it numero uno} Illingworth showed that the universe at z $\sim$ 1 looks much like the universe today. He emphasized that since most of the 
baryons reside in
ellipticals and bulges today, these largely define galaxy evolution.

\section{Conclusions}

We are coming into a new era of Galactic investigation where it is now
possible to identify fossil evidence of ancient events.  In the
Galactic halo, we might label these the Helmi event, the $\omega$~Cen
event, the Sgr dwarf event and the ``thick disk" event (i.e.\ the event
which gave rise to the thick disk). Some of these halo objects may even
be related.

In fact, there are likely to be many signatures of the galaxy formation
process, not only substructure in the outer halo (Helmi, Morrison).
Some of these, like the existence of truncation radii (van der Kruit)
and the distribution of specific angular momentum between the Galactic
components (Fall), may have been set down long ago.  In all likelihood,
we have yet to recognize many important signals, i.e.\ fossil
signatures, of past events.

One such signature may be the existence of superthin disk galaxies,
examples of which were shown in several talks.
It is interesting that superthin disk galaxies exist at all. These may be
telling us that, for the most part, disk formation is highly quiescent.
It simply putters along over 10 Gyr or so. It is tempting to suggest
that the Madau plot, which purports to monitor the star formation
history over cosmic time, is largely dominated by star formation in
merger events. In this picture, all disk thickening is due to dynamical
events like accretion.

To come back to Ken's original wish, it is becoming increasingly 
clear that ``to understand the chain of events that occurred during
the formation of the Galaxy'', the astronomical community {\sl must 
embrace the wealth of near-field information with the same relish that
it embraces the far field universe}. There are many compelling reasons
to support this statement. In particular, two key events in the early
universe are the epoch when dark matter haloes first became established
(roughly 14 Gyr ago), and the epoch when disks first established
themselves (10$-$12 Gyr ago).  The world-model parameters ($\Omega_m =
0.3$, $\Omega_\Lambda = 0.7$) place both of these events in the
redshift window which defines contemporary observational cosmology ($z
\approx 1-6$) at optical/IR wavelengths.

Many of the luminous objects we observe at high redshift are chemically
and dynamically evolved as shown by solar abundance levels (the dynamical
timescales are short in the cores of galaxies). To understand galaxy
evolution, we would like information from the outer parts of high-redshift
galaxies, which is exceedingly hard to obtain. This information
can be extracted from near-field studies although now the systems 
are viewed at a much later stage in their evolution. In any event,
near-field and far-field data are entirely complementary, and a
successful theory of galaxy formation will need to be firmly anchored
to both extremes.

\end{document}